\documentclass[12pt]{article}
\usepackage{amssymb}
\usepackage{amsmath}
\usepackage{epsfig}
\usepackage{graphicx}
\usepackage{wrapfig}

\begin{document}

\title{The annihilation
decays $B^-_c \to \eta^{'}(\eta ,\pi^0) l^-
\bar{\nu}$\footnote{Supported by National Natural Science
Foundation of China.} } \vspace{3mm}

\author{\small{ Li Gang$^{2}$, Ma Wen-Gan$^{1,2}$, Jiang Yi$^{2}$,
Zhang Ren-You$^{2}$, Han Liang$^{2}$ and Li Xue-Qian$^{3}$ }\\
{\small $^{1}$ CCAST (World Laboratory), P.O.Box 8730, Beijing 100080, P.R.China} \\
{\small $^{2}$ Department of Modern Physics, University of Science and Technology of China}\\
{\small (USTC), Hefei, Anhui 230026, P.R.China}\\
{\small $^{3}$ Department of Physics, Nankai University, Tianjin
300071, China} } \vskip 5cm

\date{}
\maketitle \vskip 12mm

\begin{abstract}
We investigate the simileptonic OZI-forbidden annihilation decays
$B^-_c \to \eta^{'}(\eta ,\pi^0) l^- \bar{\nu}$ for $l=\mu,e$ in
the perturbative QCD, and carry out a precise calculation without
any approximation for the one-loop contributions, which involves
integrals of 4- and 5-point loop functions. Our results show that
the branching ratios of decays $B^-_c \to \eta^{'} l^- \bar{\nu}$,
$B^-_c \to \eta l^- \bar{\nu}$ and $B^-_c \to \pi^0 l^-
\bar{\nu}$, turn out to be of orders $10^{-4}$, $10^{-5}$ and
$10^{-6}$, respectively, which could be observable in the future
experiments at the LHC.
\end{abstract}


\vfill \eject

\section{Introduction}
\par
Being the lowest bound state of two heavy quarks(charm and bottom)
with open (explicit) flavors, the $B_c$ meson provides a unique
window into the heavy quark physics, which has caused wide
experimental and theoretical investigations. After the report of
the CDF collaboration on the observation of the $B_c$ ground state
at the Fermilab Tevatron \cite{Fermilab}, people believe that it
is possible to accumulate more $B_c$ meson events in the
experiments at the Tevatron Run II\cite{Teva1}\cite{Teva2} and the
future CERN Large Hadron Collider(LHC). At the CERN LHC with the
luminosity of about ${\cal L} \sim 10^{34}cm^{-2}s^{-1}$ one can
expect around $5 \times 10^{10} ~B_c$ events per year\cite{LHC}.
Unlike symmetric heavy quarkonium ($c\bar{c}$ and $b\bar{b}$ bound
states), the $B_c$ meson is composed of heavy quarks with
different flavors and it lies below the $B\bar{D}$-threshold, so
its decays via strong and electromagnetic interactions are
forbidden. Therefore, the investigation of the $B_c$ meson decay
can offer special information compared to symmetric heavy
quarkonium. In the framework of the SM its decays can occur via
three mechanisms: (1) the c-quark decay with the b-quark being a
spectator, (2) the b-quark decay with the c-quark being a
spectator, (3) b-quark and c-quark annihilation. The first two
mechanisms are expected to contribute about $90\%$ of the total
width, and the remaining $10\%$ is owed to the  annihilation
process.

\par
There is another decay mode which does not belong to the three
aforementioned types, and it can only occur via the OZI processes.
As we know that the Okubo-Zweig-Iizuka (OZI) rule\cite{OZI} plays
an important role in the processes which occur via strong
interaction and in general at the parton level the concerned
calculations are carried out in the framework of the perturbative
QCD (PQCD). Thus careful studies on these OZI-forbidden processes
can deepen our insight to the perturbative QCD. But these
processes are very difficult to evaluate, because one not only
needs to carry out complicated loop calculations at the parton
level, but also has to deal with the non-perturbative effects of
the QCD, which is involved in the hadronization of partons.
Because of lack of solid knowledge on the non-perturbative QCD,
for the whole calculation one has to adopt concrete models which
may contaminate the theoretical results. The factorization scheme
is just to properly separate calculable perturbative part from the
non-perturbative contributions which must be evaluated either in
terms of experimental data or using concrete models\cite{Sterman}.
In estimating the B decays, model-dependent wave-functions are
adopted for the light mesons which are the final decay products.
Among most of the commonly used wave-functions, the light-cone
wave-function is more favorable, because the finally produced
mesons are light and so that should be more relativistic.
Obviously the errors brought out by using such model-dependent
wave-function are not accurately estimated, but in some sense they
are controllable while one uses as much as possible information
from available data. In order to reduce the theoretical
uncertainties of the non-perturbative QCD, the meson wave
functions should be well theoretically investigated and
experimentally tested. To gain more knowledge about the whole
picture, one hopes to make the part which can be calculated in the
framework of the perturbative QCD, as accurate as possible and to
employ more reasonable model-dependent wave-functions which have
been tested by fitting data obtained from precise measurements to
gain final results. Comparing the results with experimental data,
one may obtain information about both governing mechanisms which
are calculated in perturbative framework and the wave-functions.
In the case, we can also find a trace of new physics.

\par
The OZI-forbidden process $B^-_c \to \eta^{'} l^- \bar{\nu}$ was
studied by Sugamoto and Yang years ago \cite{yang}. In their work,
an effective Lagrangian was adopted to avoid introducing the
$B_{c}$ meson wave function, meanwhile they dealt with the light
meson by using an effective $g_{a}^*g_{b}^*\to \eta^{'}$ coupling
\cite{yang}\cite{Close}, which was obtained in the
NRQM(non-relativistic quark model) approximation. The valence
quark $q$ and anti-quark $\bar{q}$ in the light meson were assumed
to possess equal momenta and be on their mass shells, i.e.,
$p_q=p_{\bar{q}}$ and $p_q^2=m_q^2$. With such approximation and
kinematic assumption, they tactfully reduced the complicated
Feynman four-point and five-point integral
functions\cite{loop1,loop2} to be expressed in three-point
functions. As for the heavy meson $B_c$, the authors neglected the
relative momentum and binding energy of heavy quark constituents.
Then they obtained a reasonable branching ratios of $B^-_c \to
\eta^{'} l^- \bar{\nu}$. Furthermore they estimated the branching
ratio of $B^-_c \to \pi^0 l^- \bar{\nu}$ based on $J/\psi$ decay.

\par
In this paper, we will investigate the OZI forbidden\cite{OZI}
$B^-_c$ rare decay of $B^-_c \to \eta^{'}(\eta ,\pi^0) l^-
\bar{\nu}$ in the perturbative QCD without any effective coupling
approximation and kinematic assumption in one-loop Feynman diagram
evaluation. While calculating the amplitudes of these decay
processes, we have to properly deal with the dynamics of bound
states. As for the $B_c$ meson, we adopt the way similar to that
shown in Ref.\cite{Sterman}, namely, we ignore the relative
momentum of the two heavy constituents and their binding energy in
comparison with their masses. Furthermore, it is commonly assumed
that the constituents of $B_c$ are on mass shell and move together
with the same velocity for simplicity. In dealing with the final
light meson states, we keep an arbitrary relative momentum for the
light quarks $q$ and $\bar{q}$ and take the valence quarks
(anti-quarks) of the light mesons to be on their mass shells. Due
to the mass difference between up, down and strange quarks, the
decay widths of processes $B^-_c \to \eta^{'}(\eta ,\pi^0)l^- \nu$
would receive non-zero contributions. By this method, obviously,
the advantages of simplifying the loop integral calculations
vanish. Namely, one cannot reduce the five-point and four-point
loop integral functions into simple expressions involving
three-point integral functions. However, with the progress in the
technique of calculating loop diagrams, we can directly calculate
the five-point and four-point one-loop integrals numerically in
general case without any additional approximation and kinematic
assumption\cite{loop1}. Then we can obtain non-zero rates for the
decay modes which obviously violate isospin.

\par
The paper is organized as follows: In Section 2, we derive the
amplitudes. The numerical results of the decay rates for the
$B^-_c \rightarrow \eta^{'}(\eta ,\pi^0) l^- \bar{\nu}$ processes
are presented in Section 3, along with all the necessary
parameters being listed explicitly. Finally, the conclusion is
drawn in the last section.

\vskip 10mm
\section{Calculation}
\par
\subsection{The amplitude at parton level }

\par
Being OZI-forbidden processes, there is no contribution at the
tree-level for the decay channel $B^-_c \to P l^- \bar{\nu}$, and
at one-loop level six diagrams which contribute to the decay
width, are shown in Fig.1. Here $P$ stands for a pseudoscalar
meson $\eta^{'}(\eta ,\pi^0)$, and $l=\mu$ or $e$ respectively.
The six diagrams can be divided into three parts as Fig.1
(a,d),(b,e) and (c,f). It is clear that the figures in
Fig.1(d),(e) and (f) can be obtained by exchanging two internal
gluons of the corresponding diagrams in Fig.1(a),(b) and (c),
respectively.

\begin{figure}[!htb]
\begin{center}

\begin{tabular}{cc}
{\includegraphics[width=12cm]{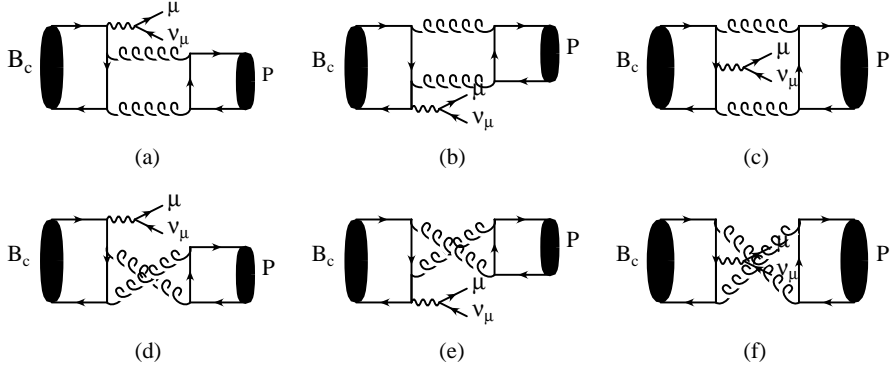}}
\end{tabular}
\caption{The diagrams at one-loop level for the process $ B^-_c
\to P l^- \nu$, where P stands for a pseudoscalar meson
$\eta^{'}(\eta ,\pi^0)$}
\end{center}
\label{fig1}
\end{figure}

\par
According to the Feyman rules, the amplitudes at parton level
corresponding to Fig.1(a), (b) and (c) can be written in the forms
as shown in Appendix A. By introducing necessary notations(see
Appendix B) and carrying out tedious, but straightforward
derivations, we obtain the explicit expressions of the amplitudes
for Fig.1(a)-(c). For Fig.1(a), it is
\begin{eqnarray}
{\cal M}_{a}  &=&-\frac{i\pi^2 g_s^4 g^2V_{cb}T^aT^b\otimes
       T^bT^a}{8[(p_1+p_2-p_3-p_4)^2-M_W^2][(p_2-p_3-p_4)^2-m_c^2]
       (2\pi)^4}\nonumber\\
       &&\times i \varepsilon_{\alpha\sigma\beta\delta}
[-m_c^2(X^{\beta\alpha\mu\rho}+i\varepsilon^{\beta\alpha\mu\rho})D^{\sigma}_a(u,m_
q)-g_{\nu\xi}(Y^{\beta\nu\alpha\theta\mu\rho}+Z^{\beta\nu\alpha\theta\mu\rho})\nonumber \\
       && \times
(p_3+p_4-p_2)_{\theta}D^{\sigma\xi}_a(u,m_q)]\bar{u}_q(p_3)\gamma^{\delta}\gamma_5
      v_q(p_4)\bar{v}_c(p_2)\gamma_{\rho}\gamma_5u_b(p_1)\nonumber\\
       && \times\bar{l}
    \gamma_{\mu}(1-\gamma_5)\nu_l.
\end{eqnarray}
The amplitude for Fig.1(b) is expressed as
\begin{eqnarray}
{\cal M}_{b} &=&-\frac{i\pi^2 g_s^4 g^2V_{cb}T^aT^b\otimes
       T^bT^a}{8[(p_1+p_2-p_3-p_4)^2-M_W^2][(p_1-p_3-p_4)^2-m_b^2]
       (2\pi)^4}\nonumber\\
       &&\times i \varepsilon_{\alpha\sigma\beta\delta}
[-m_b^2(X^{\mu\beta\alpha\rho}+\varepsilon^{\mu\beta\alpha\rho})D^{\sigma}_b(u,m_q
)-g_{\nu\theta}(Y^{\mu\xi\beta\nu\alpha\rho}+Z^{\mu\xi\beta\nu\alpha\rho})\nonumber \\
       &&\times (p_1-p_3-p_4)_{\xi}
       D^{\sigma\theta}_b(u,m_q)]\bar{u}_q(p_3)\gamma^{\delta}\gamma_5
      v_q(p_4)\bar{v}_c(p_2)\gamma_{\rho}\gamma_5u_b(p_1)\nonumber\\
       && \times\bar{l}
    \gamma_{\mu}(1-\gamma_5)\nu_l,
\end{eqnarray}
and the amplitude corresponding to Fig.1(c) reads
\begin{eqnarray}
{\cal M}_{c}  &=&-\frac{i\pi^2 g_s^4 g^2V_{cb}T^aT^b\otimes
       T^bT^a}{8[(p_1+p_2-p_3-p_4)^2-M_W^2]
       (2\pi)^4}\nonumber\\
       &&\times i \varepsilon_{\alpha\sigma\beta\delta}
[m_bm_c(X^{\beta\mu\alpha\rho}-i\varepsilon^{\beta\mu\alpha\rho})E^{\sigma}_c(u,m_
q)-g_{\nu\xi}g_{\lambda\theta}(Y^{\beta\nu\mu\lambda\alpha\rho}+Z^{\beta\nu\mu\lambda
\alpha\rho})\nonumber \\
       &&\times E^{\sigma\theta\xi}_c
       (u,m_q)] \bar{u}_q(p_3)\gamma^{\delta}\gamma_5
      v_q(p_4)\bar{v}_c(p_2)\gamma_{\rho}\gamma_5u_b(p_1)\nonumber\\
       && \times\bar{l}
    \gamma_{\mu}(1-\gamma_5)\nu_l.
\end{eqnarray}
The notations $D^{\sigma}_a(u,m_q),~D^{\sigma\xi}_a(u,m_q)$,
$D^{\sigma}_b(u,m_q)$, $D^{\sigma\theta}_b(u,m_q)$,
$E^{\sigma}_c(u,m_q)$ and $E^{\sigma\theta\xi}_c(u,m_q)$,  are
defined as the one-loop vector/tensor integrals of four- and
five-point functions\cite{loop1}\cite{loop2}, integrating over the
loop internal momentum k and their explicit expressions can be
found in Appendix B.
\par
The definitions of the variables $X$, $Y$ and $Z$ can be known
from the following identities,
\begin{eqnarray}
\gamma^{a} \gamma^{b} \gamma^{c}
&=&\gamma^{a}g^{bc}-\gamma^{b}g^{ac}+
\gamma^{c}g^{ab}-i\varepsilon^{abc\mu}\gamma_{\mu}
=(g^{a\mu}g^{bc}-g^{b\mu}g^{ac}+g^{c\mu}g^{ab})\gamma_{\mu}
-i\varepsilon^{abc\mu}\gamma_{\mu}\gamma_5\nonumber\\
   &=&X^{abc\mu}\gamma_{\mu}-i\varepsilon^{abc\mu}\gamma_{\mu}\gamma_5
\end{eqnarray}
\begin{eqnarray}
 \gamma^{a}\gamma^{b}\gamma^{c}\gamma^{d}\gamma^{e}&=&[(\varepsilon^{cde\mu}g^{ab}-
   \varepsilon^{bde\mu}g^{ac}+\varepsilon^{ade\mu}g^{bc}+\varepsilon^{abc\mu}g^{de}-
   \varepsilon^{abce}g^{d\mu}+\varepsilon^{abcd}g^{e\mu})i\gamma_5\nonumber\\
   &&+g^{ab}g^{cd}g^{e\mu}-g^{ac}g^{bd}g^{e\mu}+g^{ad}g^{bc}g^{e\mu}-
   g^{ab}g^{ce}g^{d\mu}+g^{ac}g^{be}g^{d\mu}\nonumber\\
   &&-g^{ae}g^{bc}g^{d\mu}+
   g^{ab}g^{de}g^{c\mu}-g^{ac}g^{de}g^{b\mu}+g^{bc}g^{de}g^{a\mu}-g^{a\mu}g^{be}g^{cd}\nonumber\\
   &&+g^{ae}g^{b\mu}g^{cd}+g^{a\mu}g^{bd}g^{ce}-g^{ad}g^{b\mu}g^{ce}-
   g^{ae}g^{bd}g^{c\mu}+g^{ad}g^{be}g^{c\mu}]\gamma_{\mu}\nonumber\\
   &=&Y^{abcde\mu}\gamma_5\gamma_{\mu}+Z^{abcde\mu}\gamma_{\mu}
\end{eqnarray}
The expressions of the contributions from other three
diagrams(Fig.1(d)-(f)) are similar to that of the first three (for
Fig.1(a)-(c)). For convenience, we omit their explicit expressions
in the text. It is noted that the contribution of Fig.1(c)
involves five-point tensor integration functions. We follow the
approach in Ref.\cite{loop1} to calculate directly the five-point
scalar and tensor integrals.

\par
\subsection{The hadronic matrix elements}

\par
In above subsection we derive the amplitudes corresponding to the
Feynman diagrams shown in Fig.1 at the parton level. In order to
obtain the decay rates, one has to evaluate the hadronic matrix
elements. Hadronization happens at the energy scale of
$\Lambda_{QCD}$ which is the region governed by the
non-perturbative QCD. So far, there is no any reliable way to
evaluate the hadronic matrix elements from any underlying theory.
Instead, to do this job, one needs to invoke concrete models. The
initial meson $B_c$ is composed of only heavy quarks, so that we
can suppose the two valence quarks $b$ and $\bar{c}$ in $B_c$
meson to be on mass shell approximately and the composition of
$B_c$ can be well described by its wavefunction at origin. On the
contrary, the produced pseudoscalar meson is composed of light
quark and antiquark whose three-momentum are much larger than
$\Lambda_{QCD}$, so that they are very relativistic, in this case,
the light-cone wavefunctions seem to be applicable for the
hadronization of light
mesons\cite{light-cone1,light-cone2,light-cone3}.

\par
For the pseudoscalar mesons, the SU(3) flavor wavefunctions are
\begin{equation}
\pi^{0}={u\bar{u}-d\bar{d}\over \sqrt{2}},\;
\eta_{0}={d\bar{d}+u\bar{u}+s\bar{s}\over \sqrt{3}}\;{\rm and}\;
\eta_{8}={d\bar{d}+u\bar{u}-2s\bar{s}\over \sqrt{6}},
\end{equation}
$\eta$ and $\eta^{'}$ are mixtures of $\eta_{0}$ and $\eta_{8}$,
and
\begin{equation}
\eta=\cos\theta\eta_{8}-\sin\theta\eta_{0},\;\;\;\;
\eta^{'}=\sin\theta\eta_{8}+\cos\theta\eta_{0}.
\end{equation}
Among the final produced light mesons concerned in the decays of
$B_c\rightarrow \pi^0( \eta, \eta')+l\bar \nu$, only $\eta_0$ is
SU(3) singlet. Therefore, the decays $B_c\rightarrow \pi^0+l\bar
\nu$ and $B_c\rightarrow \eta_8+l\bar \nu$ are related to isospin
or $SU_(3)$ violation. In general, there are two possibilities
which can induce isospin or SU(3) violation. The first is photon
emission and the second is due to mass splitting of u and d
quarks. In our case, the violation comes from the quark mass
splitting in the effective Hamiltonian. There are two sources
which are related to the quark mass and can contribute to the
amplitude. One is from the Feynman diagrams where the quark
propagators contain quark masses, and another is from the
higher-twist parts of the wavefunction of meson. But normally we
calculate the hadron matrix elements with only the leading-twist
part of light pseudoscalar wave function with valence quarks
on-shell, since it can give enough accuracy and simplify our
calculations.

\par
The hadronic matrix element part for process $B^-_c \to P l^- \nu$
can be written as
\begin{eqnarray}
\label{amplitude}
 &&\left<P(p_{_P})|\bar{q}\gamma_{\alpha}\gamma_{5}q \sum_i C^{\mu\alpha\beta}_i(u,m_q) \bar{c}
 \gamma_{\beta}\gamma_{5}b|B_c(p_{_{B_c}})\right>\nonumber\\
&&=-i f_{_P}p_{_P\alpha}\int_{0}^{1}du\phi_{_P}(u,\mu)\sum_i
C^{\mu\alpha\beta}_i(u,m_q) i f_{_{B_c}}p_{_{B_c\beta} },
  \end{eqnarray}
where $P$ represents $\pi^0$, $\eta$ and $\eta'$, $C_i$ are the
coefficients, the summation is over the diagrams shown in last
subsection, and in Eq.(\ref{amplitude}) we used the expression of
the leading-twist light-cone distribution amplitudes for
pseudoscalar mesons($\pi^0$, $\eta$ and $\eta'$) with flavor
content($\bar q q$) written as\cite{light-cone2}
\begin{equation}
<P(p_{_P})|\bar
q(y)_{\alpha}q(x)_{\beta}|0>_{(x-y)^2=0}=\frac{if_{_P}}{4}(\rlap/{p}_{_P}
\gamma_5)_{\beta\alpha} \int_0^1 du e^{i(\bar u
qx+uqy)}\phi_{_P}(u,\mu).
\end{equation}
where the "bar"-notation over $u$ is defined as $\bar u\equiv
1-u$, and parameter $\mu$ is the renormalization scale of the
light-cone operators on the left-hand side. Also the light-cone
wavefunction of the leading twist is normalized to unity:
\begin{equation}
\int_{0}^{1}du\phi_{_P}(u,\mu)=1,~~(P=\pi^0,\eta,\eta').
\end{equation}
The asymptotic distribution amplitude is defined as the limit in
which the renormalization scale goes to infinity. The explicit
asymptotic forms of the leading twist light-cone wavefunctions of
the light pseudoscalar mesons can be different. There are several
typical leading twist light-cone distribution amplitudes for the
light pseudoscalar mesons, so far, one cannot determine which one
is the most suitable. In our following calculation, we take three
different types of the leading twist light-cone wavefunctions for
light meson, which are frequently adopted
\cite{light-cone2,wave1,wave2,wave3}. They are expressed as
\begin{eqnarray}
&&\label{fun1} \phi_{_{P,1}}(u,\mu \to \infty) = 6u(1-u), \\
&&\label{fun2} \phi_{_{P,2}}(u,\mu \to \infty) = 30u^2(1-u)^2, \\
&&\label{fun3} \phi_{_{P,3}}(u,\mu \to \infty) = {15\over
2}(1-2u)^2[1-(1-2u)^2].
\end{eqnarray}

Finally, we obtain the hadronic matrix elements of ${\cal M}_a$,
${\cal M}_b$ and ${\cal M}_c$ as follows
\begin{eqnarray}
\langle Pl\bar{\nu}|{\cal M}_{a}|B_c\rangle &=&-\frac{i\pi^2 g_s^4
       g^2V_{cb}T^aT^b\otimes
       T^bT^a}{8(2\pi)^4N_c^2[(p_{_{B_c}}-p_{_P})^2-M_W^2][(\frac{m_c}{M_{B_c}}p_{_{B_c}}-p_{_P})^2-m_c^2]
       }\int_0^1du\phi_{_P}(u,\mu \to \infty)\nonumber\\
       &&\times \{i f_{_P} f_{_{B_c}}p_{_P}^{\delta} p_{_{B_c\rho}} \varepsilon_{\alpha\sigma\beta\delta}
       [-m_c^2(X^{\beta\alpha\mu\rho}+i\varepsilon^{\beta\alpha\mu\rho})\sum_{q=u,d,s}D^{\sigma}_a(u,m_q)\nonumber\\
       &&-g_{\nu\xi}(Y^{\beta\nu\alpha\theta\mu\rho}+Z^{\beta\nu\alpha\theta\mu\rho})\left(p_{_P}-\frac{m_c}{M_{_{B_c}}}p_{_{B_c}}\right)_{\theta}\sum_{q=u,d,s}D^{\sigma\xi}_a
       (u,m_q)]\}\nonumber\\
       && \times\bar{l}
    \gamma_{\mu}(1-\gamma_5)\nu_l,
\end{eqnarray}
\begin{eqnarray}
\langle Pl \bar{\nu}|{\cal M}_{b}|B_c\rangle &=&-\frac{i\pi^2
g_s^4 g^2V_{cb}T^aT^b\otimes
       T^bT^a}{8(2\pi)^4N_c^2[(p_{_{B_c}}-p_{_P})^2-M_W^2][(\frac{m_b}{M_{B_c}}p_{_{B_c}}-p_{_P})^2-m_b^2]
       }\int_0^1du\phi_{_P}(u,\mu \to \infty)\nonumber\\
       &&\times \{i f_P f_{_{B_c}}p_{_P}^\delta p_{_{B_c\rho}}\varepsilon_{\alpha\sigma\beta\delta}
       [-m_b^2(X^{\mu\beta\alpha\rho}+\varepsilon^{\mu\beta\alpha\rho})\sum_{q=u,d,s}D^{\sigma}_b(u,m_q)\nonumber\\
       &&-
       g_{\nu\theta}(Y^{\mu\xi\beta\nu\alpha\rho}+Z^{\mu\xi\beta\nu\alpha\rho})\left(\frac{m_b}{M_{B_c}}p_{_{B_c}}-p_{_P}\right)_{\xi}\sum_{q=u,d,s}
       D^{\sigma\theta}_b(u,m_q)]\}\nonumber\\
       && \times\bar{l}
    \gamma_{\mu}(1-\gamma_5)\nu_l,
\end{eqnarray}
\par

\begin{eqnarray}
\langle Pl \bar{\nu}|{\cal M}_{c}|B_c\rangle &=&-\frac{i\pi^2
g_s^4 g^2V_{cb}T^aT^b\otimes
       T^bT^a}{8(2\pi)^4N_c^2[(p_{_{B_c}}-p_{_P})^2-M_W^2]
       }\int_0^1du\phi_{_P}(u,\mu \to \infty)\nonumber\\
       &&\times \{i f_{_P} f_{_{B_c}}p_{_P}^\delta p_{_{B_c\rho}} \varepsilon_{\alpha\sigma\beta\delta}
       [m_bm_c(X^{\beta\mu\alpha\rho}-i\varepsilon^{\beta\mu\alpha\rho})\sum_{q=u,d,s}E^{\sigma}_c(u,m_q)\nonumber\\
       &&-
       g_{\nu\xi}g_{\lambda\theta}(Y^{\beta\nu\mu\lambda\alpha\rho}+Z^{\beta\nu\mu\lambda\alpha\rho})\sum_{q=u,d,s}E^{\sigma\theta\xi}_c
       (u,m_q)]\}\nonumber\\
       && \times\bar{l}
    \gamma_{\mu}(1-\gamma_5)\nu_{_l}.
\end{eqnarray}

Then we can get
\begin{eqnarray}
 \frac{dBr(B^-_c\to P l \nu)}{dsdt}=\frac{1}{(2\pi)^3}\frac{1}{32
 M_{_{B_c}}^3}|{\cal M}|^2 \tau_{_{B_c}}
\end{eqnarray}
where ${\cal M}$ is the total amplitude for the process $B^-_c \to
P l^- \nu$ at one-loop level and the Mandelstam variables
$s=(p_P+p_l)^2$, $t=(p_l+p_\nu)^2$.

\par
\section{Numerical results}

\par
In our calculation, no ultraviolet(UV) divergence appears, but
there is a mild superficial infrared(IR) divergence problem. Our
method to check the cancellation of the IR divergence is standard.
Namely, we assign a small gluon mass as a regulator and vary it to
check if the result is stable. In practical calculations, we set
the small gluon mass varying in the range between $10^{-4}~MeV$ to
$10^{-2}~MeV$ and find that the result changes only with a
negligible small fraction. Therefore, we can trust the obtained
result which is free of IR problem. In following numerical
calculation we set the gluon mass being $10^{-3}~MeV$.

\par
The input parameters which we are going to use in the numerical
computations are taken as follows
\cite{yang}\cite{light-cone1}\cite{zhang}\cite{data0}\cite{data1}
\cite{data2}\cite{coupling}\cite{new}:$f_{B_c}
= 500MeV$, $f_\pi = 131MeV$, $f_\eta = f_{\eta^{'}} = 157MeV$,
 $m_\pi = 134.9766MeV$, $m_\eta = 547.75MeV$,
$m_{\eta^{'}} = 957.78MeV$, $m_b = 4800MeV$, $\alpha_s(m_{B_c}) =
0.20$, $m_c = 1500MeV$, $M_{B_c} = 6300MeV$, $V_{cb} =0.04$,
$\tau_{B_c} = 0.46ps$, the mixing angle of $\eta,\; \eta'$ $\theta
= -11^\circ$, and three possible leading twist light-cone
wavefunctions of pseudoscalar light meson are given in
Eqs.(\ref{fun1},\ref{fun2},\ref{fun3}). During the calculation, we
keep an arbitrary relative momentum for the light valence
quark(antiquark) $q(\bar{q})$, and moreover, $q(\bar{q})$ is on
mass shell, while dealing with the light meson. \pagebreak[4]
\begin{table}[h]
\begin{center}
\begin{tabular}{|c|c|c|c|c|c|c|c|} \hline
Process &$m_u$&$m_d$&$m_s$&$10^6\times BR(\phi_{_{P,1}})$&$10^6\times BR(\phi_{_{P,2}})$&$10^6\times BR(\phi_{_{P,3}})$\\
\hline
& 1.5 & 4 & / & 0.92818 & 0.37502& 9.2459 \\
\cline{2-7}
& 2 & 4 & / & 0.50312 & 0.21671 & 6.2409 \\
\cline{2-7}
$B^-_c \to \pi^0 l^-\nu$ & 3 & 5 & / & 0.61314 & 0.28907 & 5.6131 \\
\cline{2-7}
& 3 & 7 & / & 1.42393& 0.55988& 10.793 \\
\cline{2-7}
& 4 & 6 & / & 0.63533 & 0.30333& 5.4237 \\
\hline
&$m_u$&$m_d$&$m_s$&$10^5\times BR(\phi_{_{P,1}})$&$10^5\times BR(\phi_{_{P,2}})$&$10^5\times BR(\phi_{_{P,3}})$\\
\cline{2-7}
& 2 & 4 & 80 & 0.39826 & 0.21839 & 6.6085\\
\cline{2-7}
& 2 & 5 & 90 & 0.41058 & 0.23082 & 6.7120\\
\cline{2-7}
$B^-_c \to \eta l^- \nu$ & 3 & 5 & 100 & 0.42533 & 0.24725& 6.7201 \\
\cline{2-7}
& 2 & 6 & 110 & 0.42835 & 0.25345& 6.96826 \\
\cline{2-7}
& 4 & 6 & 120 & 0.43766 & 0.27107 & 7.17638\\
\cline{2-7}
& 3 & 7 & 130 & 0.44001 & 0.29839 & 7.3408 \\
\hline
&$m_u$&$m_d$&$m_s$&$10^4\times BR(\phi_{_{P,1}})$&$10^4\times BR(\phi_{_{P,2}})$&$10^4\times BR(\phi_{_{P,3}})$\\
\cline{2-7}
& 2 & 4 & 80 & 0.52462 & 0.41698 & 2.3586 \\
\cline{2-7}
& 2 & 5 & 90 & 0.52065 & 0.40817& 2.3007 \\
\cline{2-7} $B^-_c \to
\eta^{'} l^- \nu$  & 3 & 5 & 100 & 0.51300 & 0.40567 & 2.2829  \\
\cline{2-7}
& 2 & 6 & 110 & 0.50178 & 0.39619 & 2.2713  \\
\cline{2-7}
& 4 & 6 & 120 & 0.49432 & 0.39048& 2.2498  \\
\cline{2-7}
& 3 & 7 & 130 & 0.49133 & 0.38767 & 2.2233\\
\hline
\end{tabular}
\end{center}
\caption{The branching ratios of the decays $B^-_c \to \pi^0 l^-
\nu$, $B^-_c \to \eta l^- \nu$ and $B^-_c \to \eta^{'} l^- \nu$ in
the rest frame of $B_c$ are listed, and the three columns
correspond to the three different leading twist light-cone
wavefunctions of the produced pseudoscalar light mesons
($\pi^0,\;\eta,\;\eta'$). The masses for the light quarks(u, d, s)
are in $MeV$. }
\end{table}

\par
We carry out the integration of the scalar and tensor four- and
five-point integral functions precisely. We adopted the FF
package\cite{van} in the calculation of two-, three- and
four-point integral functions, while the implementations of the
scalar and the tensor five-point integrals are done exactly by
using the Fortran programs as we used in our previous works on
$e^+e^- \to t \bar t H^0$ and $e^+e^- \to Z^0H^0H^0$
processes\cite{youyu,ZhangRY} by using the approach presented in
Ref.\cite{loop1}.

\par
In order to model the light hadronic effects, we use three
different types of wavefunctions for light mesons in our
calculation. We present the theoretical predictions on the decay
widths of $B^-_c \to \pi^0 l^- \nu$, $B^-_c \to \eta l^- \nu$ and
$B^-_c \to \eta^{'} l^- \nu$ in the rest frame of $B_c$,
corresponding to the three different leading twist light-cone
wavefunctions of light mesons respectively in Table 1.

\par
\section{Conclusion }

\par
In this work, we studied the OZI forbidden quark-level
sub-processes of $B^-_c \to \eta^{'}(\eta ,\pi^0) l^- \bar{\nu}$
in the framework of the perturbative QCD. For $B_c^-\to \pi^0
l^-\bar\nu$ process, within reasonable ranges of the masses of u
and d quarks, the branching ratio with the distribution amplitude
$\phi_{_{P,3}}$ can be about one order larger than other two
distribution amplitudes ($\phi_{_{P,1}},\phi_{_{P,2}}$) and can
reach the order of $10^{-5}$ with specific light quark masses. For
$B_c^-\to \eta l^-\bar\nu$ process, the branching ratios turn out
to be of the order $10^{-6}$ for distribution amplitudes
$\phi_{_{P,1}}$ and $\phi_{_{P,2}}$, and about $10^{-5}$ for
distribution amplitude $\phi_{_{P,3}}$. For $B_c^-\to \eta^{'}
l^-\bar\nu$ decay, the $\phi_{_{P,3}}$ can give the branch ratio
values of the order $10^{-4}$. These results seem to be at the
reach of future experiments at the LHC where it is expected to
produce $5\times 10^{10}$ $B_c$ meson events per year with
$\sqrt{s}=14~TeV$ and luminosity ${\cal
L}=10^{34}cm^{-2}s^{-1}$\cite{LHC}. We conclude that the decays
$B^-_c \to \eta^{'}(\eta , \pi^0) l^- \bar{\nu}$ could be
investigated experimentally at the LHC, and the study on these
decays can deepen our understanding on both perturbative and
non-perturbative QCD.

\vskip 10mm
\par
\noindent Acknowledgement: We are very grateful for the valuable
discussions with Ma Jian-Ping and Li Tong. This work was supported
in part by the National Natural Science Foundation of China and a
special fund sponsored by Chinese Academy of Sciences.

\vskip 10mm
\par
\noindent{ Appendix A}\\
\par
The amplitudes at parton level corresponding to Fig.1 (a), (b) and
(c) are explicitly expressed as
\begin{eqnarray}
{\cal M}_{a}  &=&\int \frac{d^4k}{(2 \pi)^4}\bar{u}_q(p_3)(-i
       g_sT^a\gamma_{\alpha})\frac{i}{-\rlap /k-m_q}(-i
       g_sT^b\gamma_{\beta})v_q(p_4)\bar{v}_c(p_2)(-ig_sT^b\gamma^{\beta})\nonumber\\
      &&\times \frac{i}{\rlap /p_4-\rlap /p_2-
      \rlap /k-m_c}(-i g_sT^a\gamma^{\alpha})\frac{i}{\rlap /p_3+\rlap /p_4-
      \rlap /p_2-m_c}\frac{-i
      g}{2\sqrt{2}}\gamma^{\mu}(1-\gamma_5)V_{cb}u_b(p_1)\nonumber\\&&\times\bar{l}\frac{-i
      g}{2\sqrt{2}}\gamma^{\nu}(1-\gamma_5)\nu_l\times
      \frac{-i}{(p_3+k)^2}\frac{-i}{(p_4-k)^2}\frac{-ig_{\mu\nu}}
      {(p_1+p_2-p_3-p_4)^2-M_W^2}.\nonumber
      \\&&=-g_s^4 \frac{ g^2 }{8} V_{cb}T^aT^b\otimes T^bT^a \frac{1}
     {(p_1+p_2-p_3-p_4)^2-M_W^2}\int\frac{d^4k}{(2 \pi)^4}
     \bar{u}_q(p_3)\gamma_{\alpha}(-\rlap
     /k+m_q)\gamma_{\beta}\nonumber\\
    &&\times  v_q(p_4)\bar{v}_c(p_2)\gamma^{\beta}(\rlap /p_4-\rlap /p_2-
      \rlap /k+m_c)\gamma^{\alpha}(\rlap /p_3+\rlap /p_4-
      \rlap /p_2+m_c)\gamma^{\mu}(1-\gamma_5)u_b(p_1)\bar{l}
    \gamma_{\mu}(1-\gamma_5)\nu_l\nonumber
    \\&&\times\frac{1}{(k^2-m_q^2)(p_3+k)^2(p_4-k)^2((p_2-p_4+k)^2-m_c^2)
     ((p_2-p_3-p_4)^2-m_c^2)},
\end{eqnarray}
\begin{eqnarray}
{\cal M}_{b}  &=&\int \frac{d^4k}{(2 \pi)^4}\bar{u}_q(p_3)(-i
       g_sT^a\gamma_{\alpha})\frac{i}{-\rlap /k-m_q}(-i
       g_sT^b\gamma_{\beta})v_q(p_4)\bar{v}_c(p_2)\frac{-i
      g}{2\sqrt{2}}\gamma^{\mu}(1-\gamma_5)V_{cb}\nonumber\\
      &&\times \frac{i}{\rlap /p_1-\rlap /p_3-\rlap /p_4-m_b}
      (-ig_sT^b\gamma^{\beta})\frac{i}{\rlap /p_1-\rlap /p_3-
      \rlap /k-m_b}(-i g_sT^a\gamma^{\alpha})u_b(p_1)\nonumber\\&&\times\bar{l}\frac{-i
      g}{2\sqrt{2}}\gamma^{\nu}(1-\gamma_5)\nu_l\times
      \frac{-i}{(p_3+k)^2}\frac{-i}{(p_4-k)^2}\frac{-ig_{\mu\nu}}
      {(p_1+p_2-p_3-p_4)^2-M_W^2}.\nonumber
       \\&&=-g_s^4 \frac{ g^2 }{8} V_{cb}T^aT^b\otimes T^bT^a \frac{1}
     {(p_1+p_2-p_3-p_4)^2-M_W^2}\int\frac{d^4k}{(2 \pi)^4}
     \bar{u}_q(p_3)\gamma_{\alpha}(-\rlap
     /k+m_q)\gamma_{\beta}\nonumber\\
    &&\times  v_q(p_4)\bar{v}_c(p_2)\gamma^{\mu}(1-\gamma_5)
     (\rlap /p_1-\rlap /p_3-\rlap /p_4+m_b)\gamma^{\beta}(\rlap /p_1-\rlap /p_3-
      \rlap /k+m_b)\gamma^{\alpha}u_b(p_1)\bar{l}
    \gamma_{\mu}(1-\gamma_5)\nu_l\nonumber
    \\&&\times\frac{1}{(k^2-m_q^2)(p_3+k)^2(p_4-k)^2((p_1-p_3-k)^2-m_b^2)
     ((p_1-p_3-p_4)^2-m_b^2)},
\end{eqnarray}

\begin{eqnarray}
{\cal M}_{c}  &=& \int \frac{d^4k}{(2 \pi)^4}\bar{u}_q(p_3)(-i
       g_sT^a\gamma_{\alpha})\frac{i}{-\rlap /k-m_q}(-i
       g_sT^b\gamma_{\beta})v_q(p_4)\bar{v}_c(p_2)(-i
       g_sT^b\gamma^{\beta})\nonumber\\
      &&\times\frac{i}{\rlap /p_4-\rlap /p_2-\rlap /k-m_c}\frac{-i
      g}{2\sqrt{2}}\gamma^{\mu}(1-\gamma_5)V_{cb}\frac{i}{\rlap
      /p_1-\rlap /p_3-\rlap /k-m_b}(-i g_sT^a\gamma^{\alpha})u_b(p_1)
      \nonumber\\&&\times\bar{l}\frac{-i
      g}{2\sqrt{2}}\gamma^{\nu}(1-\gamma_5)\nu_l\times
      \frac{-i}{(p_3+k)^2}\frac{-i}{(p_4-k)^2}\frac{-ig_{\mu\nu}}
      {(p_1+p_2-p_3-p_4)^2-M_W^2}.\nonumber
    \\&&=-g_s^4 \frac{ g^2 }{8} V_{cb}T^aT^b\otimes T^bT^a \frac{1}
     {(p_1+p_2-p_3-p_4)^2-M_W^2}\int\frac{d^4k}{(2 \pi)^4}
     \bar{u}_q(p_3)\gamma_{\alpha}(-\rlap
     /k+m_q)\gamma_{\beta}\nonumber\\
    &&\times  v_q(p_4)\bar{v}_c(p_2)\gamma^{\beta}(\rlap /p_4-\rlap
    /p_2-\rlap /k+m_c)\gamma^{\mu}(1-\gamma_5)(\rlap /p_1-\rlap
    /p_3-\rlap /k+m_b)\gamma^{\alpha}u_b(p_1)\bar{l}
    \gamma_{\mu}(1-\gamma_5)\nu_l\nonumber
    \\&& \times \frac{1}{(k^2-m_q^2)(p_3+k)^2(p_4-k)^2((p_1-p_3-k)^2-m_b^2)
     ((p_4-p_2-k)^2-m_c^2)}.
\end{eqnarray}

\vskip 10mm
\par
\noindent{ Appendix B}\\
\par
The notations of the four-point and five-point one-loop functions
in our text are defined as
\begin{eqnarray}
&& D^\sigma_a(u,m_q) = {1\over i\pi^2}\int d^4k {k^\sigma\over
[k^2-m_q^2](p_3+k)(p_4-k)^2[(p_2-p_4+k)^2-m_c^2]}, \\
&& D^{\sigma\xi}_a(u,m_q) = {1\over i\pi^2}\int d^4k
{k^\sigma(p_4-p_2-k)^\xi\over
[k^2-m_q^2](p_3+k)(p_4-k)^2[(p_2-p_4+k)^2-m_c^2]},\\
&&D^\sigma_b(u,m_q) = {1\over i\pi^2}\int d^4k {k^\sigma\over
[k^2-m_q^2](p_3+k)(p_4-k)^2[(p_3-p_1-k)^2-m_b^2]}, \\
&& D^{\sigma\xi}_b(u,m_q) = {1\over i\pi^2}\int d^4k
{k^\sigma(p_1-p_3-k)^\xi\over
[k^2-m_q^2](p_3+k)(p_4-k)^2[(p_1-p_3-k)^2-m_b^2]},\\
&& E^\sigma_c(u,m_q) = {1\over i\pi^2}\int d^4k {k^\sigma\over
[k^2-m_q^2](p_3+k)(p_4-k)^2[(p_4-p_2-k)^2-m_c^2][(p_1-p_3-k)-m_b^2]},\nonumber
\\
\\
&& E^{\sigma\theta\xi}_c(u,m_q) = {1\over i\pi^2}\int d^4k
{k^\sigma(p_4-p_2-k)^\xi(p_1-p_3-k)^\theta\over
[k^2-m_q^2](p_3+k)(p_4-k)^2[(p_4-p_2-k)^2-m_c^2][(p_1-p_3-k)-m_b^2]},\nonumber\\
\end{eqnarray}
with
\begin{eqnarray}
&&p_1 = {m_b\over M_{B_c}}p_{B_c},~~ p_2 = {m_c\over
M_{B_c}}p_{B_c},~~ p_3 = p_{_P} u,~~ p_4 = p_{_P} (1-u).\nonumber
\end{eqnarray}

\vskip 15mm
\par

\pagebreak[4]

\pagebreak[4]
\end{document}